\documentclass[aps,pra,reprint,superscriptaddress,showpacs]{revtex4-1}
\usepackage[latin1]{inputenc}
\usepackage{psfrag}
\usepackage{bm}
\usepackage{multirow,amssymb,amsmath,amsfonts}
\usepackage[demo]{graphicx}

\newcommand{\gdiv}{g_\textrm{div}}
\newcommand{\gdivlb}{\gdiv^{\textrm{lb}}}

\newcommand{\Ntr}{\mathcal{N}_{\textrm{dst}}}
\newcommand{\Ntrbound}{\Ntr^\textrm{ub}}
\newcommand{\Ndivlb}{\Ndiv^\textrm{lb}}
\newcommand{\Ndiv}{\mathcal{N}_{\textrm{div}}}

\newcommand{\drift}{\boldsymbol{v}}
\newcommand{\damping}{D}
\newcommand{\nonunital}{\boldsymbol{w}}
\newcommand{\unital}{N}
\newcommand{\deco}{d}
\newcommand{\decouni}{\deco_{\boldsymbol{0}}}
\newcommand{\decov}{\deco_{\boldsymbol{v}}}

\newcommand{\dynmap}{\Phi}

\newcommand{\state}{\rho}
\newcommand{\HLS}{H}
\newcommand{\trans}{\textrm{T}}
\newcommand{\omegaa}{\omega_A}
\newcommand{\gammap}{\gamma_{\textrm{p}}}
\newcommand{\gammaa}{\gamma_{\textrm{a}}}

\newcommand{\g}{g}
\newcommand{\gre}{\g^\textrm{r}}
\newcommand{\gim}{\g^\textrm{i}}
\newcommand{\f}{v_3}

\newcommand{\trel}{\tau_{\textrm{r}}}
\newcommand{\tcor}{\tau_{\textrm{c}}}

\newcommand{\gammaMMT}{\gamma^{\damping + \damping^\trans}}
\newcommand{\gammamax}{\gammaMMT_{\textrm{max}}}

\newcommand{\gammaMaxMMT}{\gamma_{\textrm{max}}^{\damping + \damping^\trans}}

\newcommand{\GammaMin}{\Gamma_\textrm{min}}
\newcommand{\gammaDeco}{\gamma^{\deco}}

\newcommand{\bfsig}{\boldsymbol{\sigma}}
\newcommand{\bflam}{\boldsymbol{\lambda}}
\newcommand{\bfdlam}{\delta\boldsymbol{\lambda}}

\newcommand{\Tr}{\textrm{Tr}}
\newcommand{\sig}{\sigma}
\newcommand{\sigx}{\sigma_x}

\newcommand{\sigz}{\sigma_z}

\newcommand{\I}{\mathbb{I}}
\newcommand{\A}{A}

\begin{document}

\title{Bounds for the divisibility-based and distinguishability-based non-Markovianity measures}
\author{H. M\"akel\"a}
\affiliation{QCD Labs, COMP Centre of Excellence, Department of Applied Physics, Aalto University,
P.O. Box 13500, FI-00076 Aalto, Finland}
\email{harri.makela@aalto.fi}

\begin{abstract}
We derive an upper bound for the distinguishability-based non-Markovianity measure of a two-level system and prove that for certain master equations the exact value of the measure achieves this bound. Furthermore, we obtain an easily calculable lower bound for the divisibility-based non-Markovianity measure of an $n$-level system. We illustrate the calculation of these bounds through examples, considering in detail the spin--boson model. We show that the differences between the two measures in the spin--boson model are caused by the drift vector that is also responsible for the nonunitality of the dynamical map. 
\end{abstract}


\maketitle

\section{Introduction}
 
Non-Markovian processes occurring in open quantum systems have become an active 
research topic during recent years~\cite{Rivas14}. The rapid pace of research and the 
lack of a unique, widely accepted definition of quantum non-Markovianity are evidenced by 
the large number of quantum non-Markovianity measures and witnesses presented in the literature.  
These are based, for example, on the divisibility of the dynamical map~\cite{Wolf08a,Rivas10}, distinguishability of quantum states~\cite{Breuer09},  quantum Fisher information~\cite{Lu10}, quantum mutual information~\cite{Luo12}, volume of dynamically accessible states~\cite{Lorenzo13}, nonunitality of the dynamical map~\cite{Liu13}, quantum channel capacity~\cite{Bylicka14}, $k$-divisibility~\cite{Chruscinski14a}, and the quantum regression theorem~\cite{LoGullo14,Guarnieri14}. 

The most commonly used non-Markovianity measures appear to be those based on the divisibility~\cite{Rivas10} and distinguishability~\cite{Breuer09}. In some special cases these give the same criterion for non-Markovianity, but it is known that in general the divisibility measure provides a more sensitive probe of non-Markovian dynamics than the distinguishability measure~\cite{Haikka11,Chruscinski11,Vacchini11,Liu13,Rivas14,Jiang13}.
The calculation of the distinguishability measure requires an optimization over pairs of initial states. 
Although this optimization can be simplified by choosing the initial pairs to be orthogonal states on the boundary of the state space~\cite{Wissmann12}, 
it nevertheless is a time-consuming task that includes the possibility of not finding the optimal (or even close to optimal) initial state pair.   
The calculation of the divisibility measure, on the other hand, does not require any optimization procedure. 
If the master equation is known, the value of the divisibility measure of an $n$-level system can be given in terms of the eigenvalues 
of an $(n^2-1)\times (n^2-1)$ matrix. The numerical evaluation of the eigenvalues is fast even for a rather large $n$, but  
their analytical calculation is in general impossible if the system has more than two levels.

In this paper, we address these issues by deriving an upper bound for the distinguishability-based measure of a two-level system 
and a lower bound for the divisibility-based measure of an $n$-level system. The upper bound can be calculated without the need to 
optimize over intial state pairs. Moreover, for some master equations the actual value of the distinguishability measure reaches the upper bound. 
The lower bound for the divisibility measure, on the other hand, can be obtained analytically regardless of the dimension of the system and provides a sufficient condition for the dynamical map to be nondivisible.

This paper is organized as follows. In Sec.~\ref{sec_ME}, we present the general form of the master equation of a two-level system and and introduce the canonical form of the master equation of an $n$-level system. In Sec.~\ref{sec_trace}, we define the distinguishability-based non-Markovianity measure, derive an upper bound for it in the two-level case, and prove that for certain master equations the exact value of the measure equals the upper bound. In Sec.~\ref{sec_divisibility}, we introduce the divisibility-based non-Markovianity measure and obtain a lower bound for it. In the case of a two-level system, the lower bound highlights the importance of the so-called drift vector in making the dynamics non-Markovian. We give examples of applications of our results in Sec.~\ref{sec_examples}, concentrating on the spin--boson model. We show that in this system the differences between the two measures can be attributed to the drift vector. Finally, conclusions are presented in Sec.~\ref{sec_conclusions}.

\section{Master equation \label{sec_ME}}
\subsection{Master equation in the Bloch vector notation}
We first consider the differential equation governing the time evolution of a two-level system. 
We write the state of the system at time $t$ as 
\begin{align}
\label{eq_Bloch_vector}
\state(t)=\frac{1}{2}\left[\I_2+\bflam^\trans(t)\cdot\bfsig\right],
\end{align}
where $\I_2$ is the identity matrix, $\bflam(t)=(\lambda_1(t),\lambda_2(t),\lambda_3(t))^\trans$ is the Bloch vector, $\bfsig=(\sig_1,\sig_2,\sig_3)$ 
is a vector consisting of the Pauli matrices, and $\trans$ denotes the transpose. 
The master equation can be written in terms of the Bloch vector as 
\begin{align}
\label{eq_ME}
\frac{d}{dt}
\boldsymbol{\lambda}(t)
= &\drift(t)+\damping(t)\boldsymbol{\lambda}(t).
\end{align}
Here $\drift(t)=(v_1(t),v_2(t),v_3(t))^\trans\in\mathbb{R}^3$  and $\damping(t)\in \mathcal{M}_3(\mathbb{R})$, where  $\mathcal{M}_3(\mathbb{R})$ is the set of all real $3\times 3$ matrices. The vector $\drift$ and matrix $\damping$ are called the drift vector and damping matrix, respectively. 
In this paper we take the initial time to be $t=0$.  
The solution of Eq.~\eqref{eq_ME} can be expressed with the help of a vector $\nonunital(t)\in\mathbb{R}^3$ and a matrix $\unital(t)\in \mathcal{M}_3(\mathbb{R})$ as  
\begin{align}
\label{eq_dyn_map}
\bflam(t)=\nonunital(t) + \unital(t)\bflam(0),
\end{align}
where the functions $\nonunital$ and $\unital$ are obtained as the solutions of the 
equations 
\begin{align}
\label{eq_nonunital_DE}
\frac{d}{dt}\nonunital(t)&=\drift(t)+\damping(t)\nonunital(t),\quad &\nonunital(0)=\boldsymbol{0},\\
\label{eq_unital_DE}
\frac{d}{dt}\unital(t)&=\damping(t)\unital(t), &\unital(0)=\I_3. 
\end{align}

The state of the system at time $t$ can be given with the help of a linear map $\dynmap_t:\mathcal{M}_2(\mathbb{C})\to \mathcal{M}_2(\mathbb{C})$. As $\{\I_2,\sig_1,\sig_2,\sig_3\}$ is a basis of $\mathcal{M}_2(\mathbb{C})$, 
$\dynmap_t$ is uniquely determined by defining its action on these matrices. Using Eq.~\eqref{eq_dyn_map}, we find that 
\begin{align}
\label{eq_dynmap_id}
&\dynmap_t(\I_2)=\I_2+\sum_{j=1}^3 \nonunital_j(t)\sig_j,\\
\label{eq_dynmap_sigma}
&\dynmap_t(\sig_i) = \sum_{j=1}^{3} \unital_{ij}(t)\sig_j,\quad i=1,2,3.
\end{align}
For the map $\dynmap_t$ to describe physically well-defined time evolution, 
it has to be trace preserving and completely positive (CP). From Eqs.~\eqref{eq_dynmap_id} and \eqref{eq_dynmap_sigma} 
it follows that $\dynmap_t$ is trace preserving. In the rest of the paper we assume that the master equation is such that $\dynmap_t$ is also CP for any $t\geq 0$. In the example including numerical studies on the spin--boson model we check for the complete positivity of $\dynmap_t$ explicitly. 
We call $\dynmap_t$ a dynamical map and denote by $\dynmap=\{\dynmap_t\ |\ t\geq 0\}$ the set of all dynamical maps. 

For later use, we define here the concept of a unital map. A map is said to be unital if it preserves the identity element. The dynamical map of a two-level system specified in Eqs.~\eqref{eq_dynmap_id} and~\eqref{eq_dynmap_sigma} is unital if and only if $\nonunital(t)=0$. Additionally, from Eq.~\eqref{eq_nonunital_DE} we see that if the drift vector $\drift(t)$ vanishes for every $t\geq 0$, then $\dynmap_t$ is unital for any $t\geq 0$. 

\subsection{Canonical form of the master equation \label{app_canonical}}
In this paper, we make use of the so-called canonical form of the master equation. In the following, we describe the canonical form only briefly and refer the reader to \cite{Hall14} for more details. We begin with the result that the master equation of an $n$-level system can be written as 
\begin{align}
\nonumber
&\frac{d}{dt}\state(t) = -i[\HLS(t),\state(t)]\\
&+\sum_{i,j=1}^{n^2-1} \deco_{ij}(t) \left(G_i\state(t)G_j-\frac{1}{2}\{G_jG_i,\state(t)\}\right),   
\end{align}
where $G_i\in\mathcal{M}_n(\mathbb{C})$ are Hermitian matrices such that $G_0=\frac{1}{\sqrt{n}}\I_n$ and 
$\Tr[G_i G_j]=\delta_{ij}$, and $\HLS(t)\in\mathcal{M}_n(\mathbb{C})$ is Hermitian. The so-called dehoherence matrix $\deco(t)$ is a Hermitian $(n^2-1)\times (n^2-1)$ matrix and can thus be diagonalized and has real eigenvalues. Denoting the diagonal matrix consisting of the 
eigenvalues of $\deco(t)$ by $\gammaDeco(t)$, we can write $\deco(t)=U(t)\gammaDeco(t)U^\dag (t)$, 
where $U(t)$ is a unitary matrix consisting of the normalized eigenvectors of $\deco(t)$. 
By defining the decoherence operators as $L_j(t) =\sum_{i=1}^{n^2-1}U_{ij}(t)G_i$,
we obtain the canonical form 
\begin{align}
\label{eq_ME_canonical}
\nonumber 
&\frac{d}{dt}\state(t) = -i[\HLS(t),\state(t)]\\
&+\sum_{i=1}^{n^2-1}\gammaDeco_i(t) \left[L_i(t)\state(t) L_i^\dag(t)
-\frac{1}{2}\left\{L_i^\dag(t)L_i(t),\state(t)\right\}\right],
\end{align}
where $\gammaDeco_i(t)$ are the eigenvalues of $\deco(t)$.

For a two-level system we define $G_i=\frac{1}{\sqrt{2}}\sig_i,i=1,2,3$. 
Assuming that the system is governed by the master equation~\eqref{eq_ME}, we find that
\begin{align}
\nonumber 
&\HLS(t) =-\frac{1}{4}\bigg\{[\damping_{23}(t)-\damping_{32}(t)]\sig_1\\
& +[\damping_{31}(t) - \damping_{13}(t)]\sig_2 + [\damping_{12}(t) - \damping_{21}(t)]\sig_3\bigg\}
\end{align}
and the decoherence matrix becomes 
\begin{align}
\label{eq_d}
\deco(t) = \frac{1}{2}\left\{\damping(t)+\damping^\trans (t) - \Tr[\damping(t)]\I_3 + \decov(t)\right\},
\end{align}
where we have defined 
\begin{align}
\decov(t) \equiv 
i\begin{pmatrix}
0 & -v_3(t) & v_2(t)\\
v_3(t) & 0 & -v_1(t)\\
-v_2(t) & v_1(t) & 0
\end{pmatrix}.
\end{align}

\section{Distinguishability-based measure \label{sec_trace}}
\subsection{Definition}
We first quantify non-Markovianity using the distinguishability based measure presented in Ref.~\cite{Breuer09}. According to the definition of this measure, Markovian dynamics either reduces or keeps unchanged the distinguishability of physical states, whereas non-Markovian dynamics increases the distinguishability. The distinguishability of two states $\state_1,\state_2$ is characterized by the trace distance of these states and non-Markovian dynamics is indicated by 
\begin{align}
\sigma(\state_{1,2};t) \equiv \frac{1}{2}\frac{d}{dt} \|\state_1(t)-\state_2(t) \|_1 > 0.
\end{align}
Here $\|\A\|_1=\Tr\sqrt{\A^\dagger\A}$ is the trace norm.
The amount of non-Markovianity accumulated in the time interval $[0,\infty)$  can be quantified by
\begin{align}
\label{eq_Ntr}
\Ntr(\dynmap) =\frac{1}{2} \max_{\substack{\state_{1,2}(0)}}\int_{0}^{\infty} dt\ \left[|\sigma(\state_{1,2};t)|+\sigma(\state_{1,2};t)\right]. 
\end{align}

\subsection{Upper bound for $\Ntr(\dynmap)$ in a two-level system}
Assume that $\state_j(t)$, $j=1,2$, is the state of a two-level system at time $t$. 
We denote the Bloch vector of $\state_j(t)$ by $\boldsymbol{\lambda}^j(t)$ and define
\begin{align}
\bfdlam(t) \equiv \bflam^1(t) - \bflam^2(t).
\end{align} 
For a Hermitian matrix $A\in\mathcal{M}_n(\mathbb{C})$ we have $\|A\|_1=\sum_{i=1}^{n}|\gamma^A_i|$, where $\gamma_i^A$ are the eigenvalues of $A$. 
The eigenvalues of $\bfdlam(t)\cdot\boldsymbol{\sigma}$ are $\pm\|\bfdlam(t)\|$, where  $\|\cdot\|$ is
the Euclidean vector norm. Consequently $\|\state_1(t)-\state_2(t)\|_1=\|\bfdlam(t)\|$. Since the time evolution of the Bloch vector difference is determined by the equation $\frac{d}{dt}\bfdlam(t)=\damping(t)\bfdlam(t)$, we obtain 
\begin{align}
\sigma(\state_{1,2};t) 
\label{eq_sigma_1}
&=\frac{\delta\bflam(t)^\trans [\damping(t)+\damping^\trans(t)]\delta\bflam(t)}{4\|\delta\bflam(t)\|}.
\end{align}

For any symmetric matrix $\A\in\mathcal{M}_n(\mathbb{R})$ and a  vector $\boldsymbol{x}\in\mathbb{R}^n$ 
we have the inequality $\boldsymbol{x}^\trans \A\boldsymbol{x}\leq \gamma_{\textrm{max}}^{\A}\boldsymbol{x}^\trans \boldsymbol{x}$, 
where $\gamma_{\textrm{max}}^{\A}$ is the largest eigenvalue of $\A$. The equality holds if $\boldsymbol{x}$ 
is an eigenvetor corresponding to the largest eigenvalue of $A$. We thus have the upper bound 
\begin{align}
\label{eq_sigma_2}
\sigma(\state_{1,2};t)&\leq \frac{1}{4}\gammaMaxMMT(t)\|\delta\bflam(t)\|. 
\end{align}
As has been pointed out in Ref.~\cite{Hall14}, a necessary and sufficient condition for $\Ntr(\dynmap)$ to be equal to zero is that 
\begin{align}
\label{eq_gammaMaxMTM}
\gammaMaxMMT(t) \leq 0
\end{align}
for any $t\geq 0$. 

An upper bound for Eq.~\eqref{eq_sigma_2} can be obtained  
by using the inequality $\|A\boldsymbol{x}\|\leq \|A\|\|\boldsymbol{x}\|$, where the matrix norm 
of $A\in\mathcal{M}_n(\mathbb{R})$ is defined as 
$\|A\|\equiv\max_{\|\boldsymbol{x}\|=1} \|A\boldsymbol{x}\|=\sqrt{\gamma_{\textrm{max}}^{A^\trans A}}$. 
Because $\bfdlam(t)=\unital(t)\bfdlam(0)$ and $\|\bfdlam(0)\|\leq 2$, we find that 
\begin{align}
\label{eq_sigma_3}
\sigma(\state_{1,2};t) \leq &\frac{1}{2}\gammaMaxMMT(t)\|\unital(t)\|.
\end{align}
Hence we have the inequality $\Ntr(\dynmap)\leq\Ntrbound(\dynmap)$ with $\Ntrbound(\dynmap)$ defined as
\begin{align}
\label{eq_Ntr_upper_bound}
\Ntrbound(\dynmap)\equiv\frac{1}{4}\int_{0}^{\infty} \!\!\!\!\!
dt\ [|\gammaMaxMMT(t)|+\gammaMaxMMT(t)] \|\unital(t)\|.
\end{align}

\subsection{Analytical calculation of $\Ntr(\dynmap)$ in a two-level system}
We show next that if the damping matrix fulfills certain conditions, it is possible to calculate the 
value of the distinguishability measure exactly without a need to maximize over 
the initial state pairs. The idea is to show that in some cases it is possible to find a $\bfdlam(0)$ for which the corresponding value of $\sigma$ equals the upper bound given in Eq.~\eqref{eq_sigma_3} for any $t$ for which $\sigma$ is positive, indicating that $\Ntr(\dynmap)=\Ntrbound(\dynmap)$. 

We begin by writing Eq.~ \eqref{eq_sigma_1} in an alternative form as
\begin{align}
\label{eq_sigma_4}
\sigma(\state_{1,2};t) &= \frac{\bfdlam^\trans(0)\frac{d}{dt}[\unital^\trans(t)\unital(t)]\bfdlam(0)}
{4\|\bfdlam(t)\|}.
\end{align}
Assume that 
\begin{align}
\label{eq_cond_1}
\textrm{(i)}\quad &[\damping(t),\damping(t')]=[\damping(t),\damping^\trans(t')]=0 \textrm{ and} \\
\label{eq_cond_2}
\textrm{(ii)}\quad &\damping_{ij}(t)=-\damping_{ji}(t),\quad  i\neq j,
\end{align}
for any $t,t'\geq 0$. From Eq.~\eqref{eq_cond_1} it follows that the solution of Eq.~\eqref{eq_unital_DE} can be written as $\unital(t) =\exp\left(\int_{0}^{t} ds\ \damping(s)\right)$ and that the product $\unital^\trans(t)\unital(t)$ becomes 
\begin{align}
\unital^\trans(t)\unital(t) = \exp\left(\int_{0}^{t} ds\left[\damping(s) + \damping^\trans(s) \right]\right).
\end{align}
Condition (ii) implies that $\damping(t)+\damping^\trans(t)$ is diagonal and hence 
\begin{align}
\label{eq_NTN_diag}
\unital^\trans(t)\unital(t)=
\begin{pmatrix}
e^{-2\Gamma_1(t)} & 0 & 0\\
0 & e^{-2\Gamma_2(t)} & 0\\
0 & 0 &e^{-2\Gamma_3(t)}
\end{pmatrix},
\end{align} 
where 
\begin{align}
\label{eq_Gamma_def}
\Gamma_i(t)\equiv -\frac{1}{2}\int_{0}^{t} ds\ \gammaMMT_i(s), \quad i=1,2,3.
\end{align}
Choosing $\bfdlam(0)=2(\delta_{1k},\delta_{2k},\delta_{3k})$, where $k\in\{1,2,3\}$ and $\delta_{ij}$ is the Kronecker delta, and  
using Eqs.~\eqref{eq_sigma_4},\eqref{eq_NTN_diag}, and \eqref{eq_Gamma_def} we find that
\begin{align}
\sigma(\state_{1,2};t)=\frac{1}{2}\gammaMMT_k(t) e^{-\Gamma_k(t)}.
\end{align}
We define $\GammaMin(t)\equiv\min\{\Gamma_i(t)\}$, so that $\|\unital(t)\|=e^{-\GammaMin(t)}$. Assume finally that for a fixed $k\in\{1,2,3\}$ we have 
\begin{align}
\label{eq_cond_3}
\textrm{(iii)}\quad &\gammaMaxMMT(t)=\gammaMMT_k(t) \textrm{ and }\GammaMin(t)=\Gamma_k(t)
\end{align}
whenever $\gammaMaxMMT(t)>0$. Then $\sigma(\state_{1,2};t)=\frac{1}{2}\gammaMaxMMT(t)\|\unital(t)\|$
for any $t\geq 0$ for which $\gammaMaxMMT(t)>0$. 
This expression for $\sigma$ is equal to the right hand side of Eq.~\eqref{eq_sigma_3}, implying that 
under the conditions (i)-(iii) we have $\Ntr(\dynmap)=\Ntrbound(\dynmap)$.
Furthermore, by denoting the time intervals during which $\gammaMaxMMT(t)$ is larger than zero by $(a_i,b_i),i=1,2,\ldots$, 
we obtain
\begin{align}
\label{eq_Ntr_ana}
\Ntr(\dynmap)&=\sum_{i} \left(e^{-\GammaMin(b_i)} -  e^{-\GammaMin(a_i)}\right)\\
&= \sum_i\left[\|\unital(b_i)\| - \|\unital(a_i)\|\right].
\end{align}
During the time intervals $(a_i,b_i)$ the operator norm $\|\unital(t)\|$ grows. 

If Eq.~\eqref{eq_cond_3} holds for a single index $k$, then the initial state pair maximizing the amount of non-Markovianity corresponds to  
the state pair $\state_{1,2}(0)=\frac{1}{2}(\I_2 \pm \sig_k)$. 
If it holds for two indices $k,l$, the pair can be chosen as $\state_{1,2}(0)=\frac{1}{2}[\I_2 \pm (\cos\theta\sig_k+\sin\theta\sig_l)]$, 
where $\theta\in [0,2\pi)$ is arbitrary.

Although it may be possible to extend the approach used here to a system with three or more energy levels, 
this is not straightforward. While the Bloch vector representation can be generalized to a system with any number of levels 
\cite{Bruning12}, the equation $\|\state_1(t)-\state_2(t)\|_1=\|\bfdlam(t)\|$ is not valid   
if the number of levels is higher than $2$. This equation can be made to hold for an any-dimensional system 
if the distance between quantum states is defined using the Hilbert-Schmidt norm instead of the trace norm. The Hilbert-Schmidt norm of $A\in\mathcal{M}_n(\mathbb{C})$ is defined as $\|A\|_{\textrm{HS}} =\sqrt{\Tr[A^\dagger A]}$. Unfortunately, as has been pointed out in Ref.~\cite{Laine10}, the Hilbert-Schmidt norm is not suitable for the definition of non-Markovianity as the distance between two states can increase under a trace preserving CP map if this 
norm is used.

\section{Divisibility-based measure \label{sec_divisibility}}
\subsection{Definition}
An alternative way to define non-Markovian dynamics is based on the divisibility of the dynamical map $\dynmap_t$~\cite{Wolf08a,Rivas10}. If the dynamical map 
of an $n$-level system can be written as $\dynmap_{t}=\dynmap_{t,s}\dynmap_{s}$, where $\dynmap_{t,s}$ is a linear, trace preserving, and CP  map for any $0\leq s\leq t$, then $\dynmap_t$ is said to be divisible. Non-Markovian dynamics is identified with nondivisibility and it is witnessed by a non-negative function $\gdiv$ defined as~\cite{Rivas10} 
\begin{align}
\gdiv(t) =\lim_{\epsilon\to 0}\frac{\|[\dynmap_{t+\epsilon,t}\otimes \I_n](|\phi\rangle\langle\phi|)\|_1 -1}{\epsilon},
\end{align}
where $\phi$ is the maximally entangled state $\phi\equiv\frac{1}{\sqrt{n}}\sum_{i=1}^{n}|i\rangle\otimes |i\rangle$. 
Here $\{|1\rangle,|2\rangle,\ldots,|n\rangle\}$ is an orthonormal basis of the $n$-level system. 
Non-Markovian dynamics at time $t$ is equivalent with $\gdiv(t)$ 
being positive, whereas Markovian dynamics corresponds to $\gdiv(t)=0$.  
The total amount of non-Markovianity occurring in the time interval $[0,\infty)$ can be quantified by 
the non-Markovianity measure defined as \cite{Rivas10}
\begin{align} 
\label{eq_Ndiv}
\Ndiv(\dynmap)=\int_0^{\infty} dt\, \gdiv(t). 
\end{align}
In Ref.~\cite{Hall14} it has been shown that the function $\gdiv$ can be obtained in terms of the eigenvalues 
of the decoherence matrix $\deco(t)$: For an $n$-level system $\gdiv(t)$ becomes 
\begin{align}
\label{eq_gdiv}
\gdiv(t) &= \frac{1}{2}\sum_{i=1}^{n^2-1}[|\gammaDeco_i(t)|-\gammaDeco_i(t)].
\end{align}
By noting that $\Tr[d(t)]=\sum_{i=1}^{n^2-1}\gammaDeco_i(t)$ 
and $\|\deco(t)\|_1=\sum_{i=1}^{n^2-1}|\gammaDeco_i(t)|$, this can alternatively be written as
\begin{align}
\label{eq_gdiv_2}
\gdiv(t) &= \frac{1}{2}\left\{\|\deco(t)\|_1 - \Tr[\deco(t)]\right\}. 
\end{align}
Clearly the dynamical map is nondivisible if $\Tr[\deco(t)]<0$. 
\subsection{Lower bound for $\Ndiv(\dynmap)$ in an $n$-level system}
The value of $\gdiv(t)$ is determined by the eigenvalues of $\deco(t)$. 
The analytical calculation of these is typically impossible if $n$ is larger than two. 
Hence, it would be helpful to have a way to estimate the value of $\gdiv$ without the need to know the eigenvalues of the coherence matrix. This turns out to be possible using the fact that the trace and Hilbert-Schmidt norms are related by the inequality 
\begin{align}
\label{eq_norm_equivalence}
\|A\|_{\textrm{HS}}\leq \|A\|_1.
\end{align}
With the help of Eqs.~\eqref{eq_gdiv_2} and \eqref{eq_norm_equivalence} and the Hermiticity of $\deco(t)$ we obtain the following lower bound for $\gdiv(t)$ 
\begin{align}
\label{eq_gdiv_lower_bound_general}
\gdiv(t)\geq \gdivlb(t)\equiv \frac{1}{2}\left\{\sqrt{\Tr[\deco(t)^2]} - \Tr[\deco(t)]\right\}. 
\end{align}
The corresponding non-Markovianity measure is defined as
\begin{align}
\Ndivlb(\dynmap)=\int_{0}^{\infty} dt\ \gdivlb(t).
\end{align}

\subsection{Lower bound for $\Ndiv(\dynmap)$ in a two-level system}
In the case of a two-level system the expression \eqref{eq_gdiv_lower_bound_general} can be simplified. 
We denote by $\decouni(t)$ the decoherence matrix obtained by setting the drift vector $\drift(t)$ equal to zero in Eq.~\eqref{eq_d},
\begin{align}
\nonumber 
\decouni(t)&=\frac{1}{2}\left\{\damping(t)+\damping^\trans(t)-\Tr[\damping(t)]\I_3\right\}.
\end{align}
A direct calculation gives that $\Tr[\deco(t)^2] =\Tr[\decouni(t)^2]+\frac{1}{2}\|\drift(t)\|^2$.  
Because $\Tr[\deco(t)]=\Tr[\decouni(t)]$, the lower bound of Eq.~\eqref{eq_gdiv_lower_bound_general} becomes now
\begin{align}
\label{eq_lower_bound_two_level}
&\gdivlb(t)= \frac{1}{2}\left\{\sqrt{\Tr[\decouni(t)^2]+\frac{1}{2}\|\drift(t)\|^2}-\Tr[\decouni(t)]\right\}\\
\label{eq_lower_bound_two_level_2}
&=\frac{1}{4}\bigg\{\sqrt{2\Tr[\damping^2]+2\Tr[\damping\damping^\trans]-\Tr[\damping]^2+2\|\drift\|^2}+\Tr[\damping] \bigg\},
\end{align}
where in the lower equation we do not show the time argument. If $\Tr[\damping(t)]>0$, the nondivisibility of the dynamical map is guaranteed regardless 
of the value of $\|\drift(t)\|$. If $\Tr[\damping(t)]\leq 0$, the drift vector plays an important role in making the dynamical map nondivisible: 
A sufficient condition for the nondivisibility is that
\begin{align}
\label{eq_gdiv_condition}
\|\drift(t)\|^2 > \Tr[\damping(t)]^2-\Tr[\damping(t)^2]-\Tr[\damping(t)\damping(t)^\trans].
\end{align}
Note that the requirement of the complete positivity of the dynamical map may impose  
conditions on the allowed values of $\|\drift(t)\|$. 

\section{Examples \label{sec_examples}}
In this section we show examples of the calculation of the bounds of the non-Markovianity measures in the 
case of three commonly used master equations. These are the phase-damping, amplitude-damping, and 
spin--boson master equations.

\subsection{Phase damping}
The phase damping of a two-level system is described by the master equation
\begin{align}
\frac{d}{dt}\state(t)=\gammap(t)\left[\sig_3\state(t)\sig_3 -\state(t)\right].
\end{align}
For this equation the drift vector vanishes and the damping matrix reads 
\begin{align}
\label{eq_damping_phase}
&\damping(t)=
-2\begin{pmatrix}
\gammap(t) & 0 & 0\\
0 & \gammap(t) & 0\\
0 & 0 & 0
\end{pmatrix}.
\end{align}
The eigenvalues of $\damping(t)+\damping(t)^\trans$ are $\gammaMMT_1(t)=\gammaMMT_2(t)=-4\gammap(t)$ 
and $\gammaMMT_3(t)=0$. Consequently $\gammaMaxMMT(t)=-4\gammap(t)$ whenever 
$\gammaMaxMMT(t)>0$ and $\GammaMin(t)=-\frac{1}{2}\int_{0}^{t}ds\ \gammaMMT_3(s)=0$. From the latter equation 
if follows that $\|\unital(t)\|=1$ and hence the upper bound of the distinguishability measure 
becomes $\Ntrbound=\int_{0}^{\infty} dt[|\gammap(t)|-\gammap(t)]$. 

The lower bound for $\gdiv(t)$ reads $\gdivlb(t)= |\gammap(t)|-\gammap(t)$, 
which in this case equals the exact value of $\gdiv(t)$. It follows that $\Ndivlb(\dynmap)=\Ndiv(\dynmap)$.

\subsection{Amplitude damping}
The master equation characterizing amplitude damping is
\begin{align}
&\frac{d}{dt}\state(t) =\gammaa(t)\left[\sig_-\state(t)\sig_+ - \frac{1}{2}\{\sig_+\sig_-,\state(t)\}\right],
\end{align}
where $\sig_\pm =\frac{1}{2}(\sig_1\pm i\sig_2)$.
The damping matrix and drift vector are
\begin{align}
&\damping(t) =
-\begin{pmatrix}
\frac{\gammaa(t)}{2} & 0 & 0\\
0 & \frac{\gammaa(t)}{2} & 0\\
0 & 0 & \gammaa(t)
\end{pmatrix},\quad \drift(t)
=\begin{pmatrix}
0\\
0\\
-\gammaa(t)
\end{pmatrix}.
\end{align}
It is easy to see that Eqs.~\eqref{eq_cond_1} and \eqref{eq_cond_2} hold. Because  
$\gammaMMT_1(t)=\gammaMMT_2(t)=\frac{1}{2}\gammaMMT_3(t)=-\gammaa(t)$, we have $\gammaMaxMMT(t)=\gammaMMT_3(t)$ and $\GammaMin(t)=\Gamma_3(t)$ if
$\gammaMaxMMT(t)>0$, implying that the condition (iii) in Eq.~\eqref{eq_cond_3} is fulfilled and $\Ntr(\dynmap)$ is given by Eq.~\eqref{eq_Ntr_ana} with  $\GammaMin(t)=\int_{0}^{t}ds\ \gammaa(s)$. The initial state pair yielding the maximal amount of non-Markovianity is $\state_{1,2}(0)=\frac{1}{2}(\I_2\pm \sig_3)$, as has been suggested by many authors~\cite{Breuer09,He11,Makela13}.  

A direct calculation utilizing Eq.~\eqref{eq_lower_bound_two_level} yields the lower bound $\gdivlb(t)= \frac{1}{2}[|\gammaa(t)|-\gammaa(t)]$. Similarly to  the case of the phase damping model, this lower bound equals the exact value $\gdiv(t)$ and hence $\Ndivlb(\dynmap)=\Ndiv(\dynmap)$. 

\subsection{Spin--boson model}
\label{sec_Hamiltonian}
\subsubsection{Hamiltonian}
As the last example we consider a two-level atom with  energy level separation $\omegaa$ coupled to 
an environment consisting of harmonic oscillators. The total Schr\"odinger picture Hamiltonian reads $(\hbar =1)$
\begin{align}
\nonumber 
H_{\textrm{sb}}&=H_S+H_E+H_I\\
&=\frac{\omegaa}{2}\sigz + \sum_m\omega_m \hat{b}_m^\dagger \hat{b}_m
+\sigx\sum_m (g_m\hat{b}_m + g_m^*\hat{b}^\dagger_m),
\end{align}
where $H_S,H_E$, and $H_I$ are the system, environment, and interaction Hamiltonians, respectively, the asterisk indicates the complex conjugate, the index $m$ labels the modes of the environment, $\omega_m$ is the frequency of the $m$th oscillator, $g_m$ is a mode-dependent coupling constant, and $[\hat{b}_m,\hat{b}_{l}^{\dag}] = \delta_{ml}$. The non-Markovian dynamics occurring in this system has been previously studied in Refs.~\cite{Clos12,Zeng12,Makela13}. We assume that the interaction between the open system and the environment is weak and that the initial state of the total system factorizes as $\state(0)\otimes\state_E$, where $\state_E$ is the initial state of the environment. 

In the case of a weak system-environment interaction, the master equation can be written as in Eq.~\eqref{eq_ME} with the drift vector and damping matrix defined as (see \cite{Clos12,Makela13}) 
\begin{align}
\label{eq_drift_sb}
&\drift(t) =
\begin{pmatrix}
0\\
0\\
\f(t)
\end{pmatrix},
\end{align}
and 
\begin{align}
\label{eq_damping_sb}
&\damping(t)=\begin{pmatrix}
0 & -\omegaa & 0\\
\omegaa-2\gim(t) & -2\gre(t) & 0\\
0 & 0 & -2\gre(t)
\end{pmatrix}.
\end{align}
In these equations the superscripts r and i denote the real and imaginary part, respectively, and 
the functions $\f$ and $\g$ are defined as 
\begin{align}
\g(t) &= 2\int_{0}^{t} ds \ e^{-i\omegaa s}K_1(s),\\
\f(t) 
&= -4\int_{0}^{t} ds \sin(\omegaa s)K(s),
\end{align}
where
\begin{align} 
\label{eq_K_1}
K_1(s)&=\sum_m |g_m|^2 (1+2\textrm{Tr}[\hat{b}^\dag_m\hat{b}_m\state_E])\cos(\omega_m s),\\
\label{eq_K}
K(s)&=\sum_m |g_m|^2 \sin(\omega_m s).
\end{align}
Typically $K_1$ and $K$ are referred to as the noise and dissipation kernel, respectively \cite{Breuer}. 

\subsubsection{Distinguishability measure}
Conditions (i) and (ii) given in Eqs.~\eqref{eq_cond_1} and \eqref{eq_cond_2} are not valid for the damping matrix 
of Eq.~\eqref{eq_damping_sb}, and hence the value of $\Ntr(\dynmap)$ cannot be determined exactly using Eq.~\eqref{eq_Ntr_ana}. Instead, we calculate the upper bound $\Ntrbound(\dynmap)$ and compare it to the numerically obtained value $\Ntr(\dynmap)$. We assume that the environment is in the vacuum state  
corresponding to the thermal equilibrium state at zero temperature and make the replacement $\sum_m |g_m|^2\to \int_0^\infty d\omega J(\omega)$ in Eqs.~\eqref{eq_K_1} and \eqref{eq_K}. Here $J$ is the Ohmic spectral density with a Lorentz-Drude cutoff, 
\begin{align}
J(\omega)=\frac{\alpha}{\pi}\frac{\omega}{\omegaa}\frac{\Omega^2}{\Omega^2+\omega^2},
\end{align}
where $\alpha$ characterizes the strength of the system-environment coupling and  
$\Omega$ is the cutoff-frequency. It defines the system correlation time as $\tcor=1/\Omega$. 
The relaxation time is $\trel = 1/\gre(\infty)$. We have checked numerically using the approach described in Ref.~\cite{Makela13} that the dynamical map is CP for the parameter values used here. 

\begin{figure}
\includegraphics[scale=1.1]{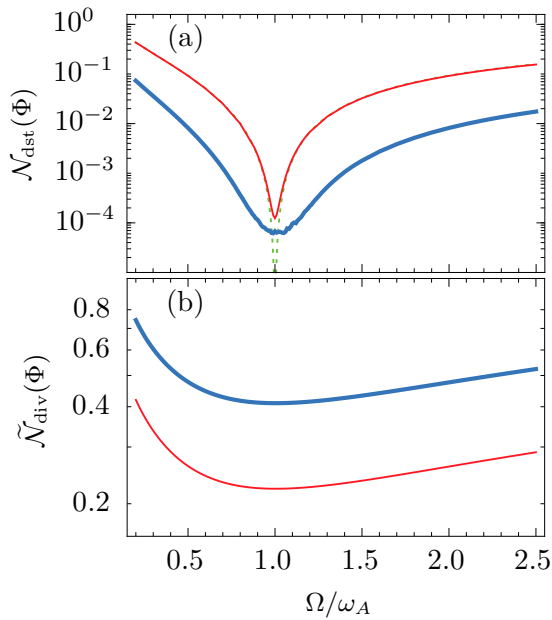}
\caption{\label{fig_N} (Color online). (a) The distinguishability measure for the spin--boson model (thick blue line), its upper bound given by Eq.~\eqref{eq_Ntr_bound_sb} (thin red line), and the approximate value of the bound shown in Eq.~\eqref{eq_Ntr_bound_sb_app} (thin green dashed line). (b) The modified divisibility measure defined in Eq.~\eqref{eq_Ndiv_alt} (thick blue line) and its lower bound given in Eq.~\eqref{eq_Ndiv_alt_lb} (thin red line). Note that in both figures the scale of the vertical axis is logarithmic. The strength of the system-environment interaction is $\alpha/\omegaa=0.01$.
}
\end{figure}

The largest eigenvalue of $\damping(t)+\damping^\trans(t)$ reads 
\begin{align}
\gammamax(t) = 2[|\g(t)| - \gre(t)]
\end{align}
and thus the upper bound for the distinguishability measure becomes 
\begin{align}
\label{eq_Ntr_bound_sb}
\Ntrbound(\dynmap)=\int_{0}^{\infty}dt\ [|\g(t)| - \gre(t)] \|\unital(t)\|.
\end{align}
The system is Markovian, $\Ntr(\dynmap)=0$, if and only if 
\begin{align}
\label{eq_distinguishability_conditions_sb}
\gim(t)=0 \textrm{ and } \gre(t)\geq 0 
\end{align}
for every $t\geq 0$. To obtain an analytical estimate for $\Ntrbound(\dynmap)$ we use the approximate expression  
$\|\unital(t)\|\approx e^{-t/\trel}$. Since in the weakly interacting system studied here the relaxation time is much longer than the correlation time, we replace $\g(t)$ with $\g(\infty)$ in Eq.~\eqref{eq_Ntr_bound_sb}, finding that  
\begin{align}
\label{eq_Ntr_bound_sb_app}
\Ntrbound(\dynmap)\approx 
\sqrt{1+\nu^2}-1,
\end{align}
where $\nu=\frac{\gim(\infty)}{\gre(\infty)}$. The dynamics is (nearly) Markovian if $\nu=0$. 
A similar result has been obtained using an alternative approach in Ref.~\cite{Makela13}. 
In Fig.~\ref{fig_N}(a), we plot $\Ntr(\dynmap)$ together with the upper bound $\Ntrbound(\dynmap)$ 
and its approximate expression given in Eq.~\eqref{eq_Ntr_bound_sb_app}. The location of the minimum, 
as well as the overall behavior of the measure, is quite well estimated by $\Ntrbound(\dynmap)$.  
Furthermore, the relative error between the exact and analytical expressions for the upper bound is small everywhere else except near the region where the non-Markovianity is very small. 

\subsubsection{Divisibility measure}
Direct calculation using Eqs.~\eqref{eq_lower_bound_two_level_2}, \eqref{eq_drift_sb}, and~\eqref{eq_damping_sb} yields the lower bound for $\gdiv(t)$ as 
\begin{align}
\gdivlb(t)= -\gre(t)+\sqrt{|\g(t)|^2-\frac{1}{2}\gim(t)^2 + \frac{1}{8}\f(t)^2}.
\end{align}
Necessary conditions for the dynamical map to be divisible are that
\begin{align}
\label{eq_divisibility_conditions_sb}
\f(t)=\gim(t)=0 \textrm{ and } \gre(t)\geq 0.
\end{align}
The exact value of $\gdiv(t)$ can be obtained in terms of the eigenvalues of the decoherence matrix. 
These are  
\begin{align}
\label{eq_gamma_i_sb}
&\gammaDeco_{1,2}(t) =\gre(t)\mp \sqrt{|\g(t)|^2+\frac{1}{4}\f(t)^2},
\end{align}
and $\gammaDeco_3(t)=0$. Clearly always $\gammaDeco_1(t)\leq 0$ and $\gammaDeco_2(t)\geq 0$, so that 
\begin{align}
\label{eq_gdiv_t_sb}
\gdiv(t)=-\gre(t)+\sqrt{|\g(t)|^2+\frac{1}{4}\f(t)^2}.
\end{align}
Hence the necessary conditions for the divisibility given in Eq.~\eqref{eq_divisibility_conditions_sb} 
are also sufficient conditions. In all three examples considered in this paper, the 
lower bound $\gdivlb(t)$ has given exactly the same conditions for the divisibility of the dynamical map as the exact value $\gdiv(t)$. 

Comparing Eqs.~\eqref{eq_distinguishability_conditions_sb} and~\eqref{eq_divisibility_conditions_sb} we observe that the differences between the two non-Markovianity measures are caused by the drift vector. Assuming that $\f(t)=0$ for any $t\geq 0$, we have the implications $\Ndiv(\dynmap)=0\Leftrightarrow \Ntr(\dynmap)=0$ [or equivalently $\Ndiv(\dynmap)>0\Leftrightarrow \Ntr(\dynmap)>0$]. Note that the condition that the drift vector vanishes for any $t\geq 0$ is a necessary and sufficient condition for $\dynmap_t$ to be unital for any $t\geq 0$. Hence the distinguishability and divisibility measures give an identical condition for non-Markovianity in the spin--boson model if the dynamical map $\dynmap_t$ is unital for any $t$. However, typically the drift vector 
is nonzero at some point of time and the two measures are not equivalent.

For the Ohmic spectral density $\gdiv(t)$ is positive if $t>0$, indicating that the system is non-Markovian for an infinitely long time interval. Similar phenomenon, referred to as eternal recoherence, has been observed in an other model in Ref.~\cite{Hall14}. Since the limit $\lim_{t\to\infty}\gdiv(t)$ is now  finite, the divisibility measure diverges. One possible way to make the measure finite is to define a modified divisibility measure as 
\begin{align} 
\label{eq_Ndiv_alt}
\widetilde{\mathcal{N}}_{\textrm{div}}(\dynmap)=\int_0^{\infty} dt\, \gdiv(t) \|\unital(t)\|. 
\end{align}
The norm of $\unital(t)$ acts as a suppressing factor yielding the integral finite.
A lower bound for $\widetilde{\mathcal{N}}_{\textrm{div}}(\dynmap)$ can be obtained by replacing $\gdiv(t)$ with $\gdivlb(t)$ in this equation,
\begin{align}
\label{eq_Ndiv_alt_lb}
\widetilde{\mathcal{N}}_{\textrm{div}}^{\textrm{lb}}(\dynmap)=\int_0^{\infty} dt\, \gdivlb(t) \|\unital(t)\|. 
\end{align}
 We show $\widetilde{\mathcal{N}}_{\textrm{div}}(\dynmap)$ and $\widetilde{\mathcal{N}}_{\textrm{div}}^{\textrm{lb}}(\dynmap)$ in Fig.~\ref{fig_N}(b). The behavior of the exact value is seen to be well estimated by the lower bound.

\section{Conclusions \label{sec_conclusions}}
In this paper, we have studied the properties of the distinguishability and divisibility-based 
non-Markovianity measures.  We have derived an upper bound for the distinguishability-based non-Markovianity measure of an arbitrary two-level system. This bound is directly obtained from the dynamical map of the system and does not require any optimization procedure. We found that for master equations fulfilling certain conditions, the exact value of the measure equals the upper bound and the initial state pair yielding the maximal non-Markovianity is easily identified. 

Similarly, we obtained a lower bound for the divisibility measure of an $n$-dimensional system. 
Unlike the exact value of the measure, this lower bound can be calculated without the need to know the eigenvalues of the decoherence matrix and hence provides a convenient analytical tool to study the divisibility-based non-Markovianity measure. This is particularly useful if the dimension of the system is higher than $2$. 

We calculated these bounds in the context of the phase- and amplitude-damping master equations and the spin--boson model. In all these systems, the lower bound and the exact expression for the divisibility measure provided identical conditions for the divisibility of the dynamical map. 
Furthermore, we found that for the amplitude-damping master equation, the upper bound for the distinguishability measure and the lower bound for the
divisibility measure are equal to the exact values of these measures. In the case of the spin--boson model, the upper and lower bounds estimate the behavior of the exact values well and the differences between the two measures are related to the nonunital character of the dynamical map.

\acknowledgments
The author thanks E.-M.~Laine for helpful discussions and M.~M\"ott\"onen for careful reading of the manuscript. This research has been supported by the Emil Aaltonen Foundation and the Academy of Finland through its Centres of Excellence Program (Project No.~251748) and Grants No.~138903 and No.~272806.

\appendix


\begin{thebibliography}{33}

\bibitem{Rivas14} \'A. Rivas, S. F. Huelga, and M. B. Plenio, Rep. Prog. Phys. {\bf 77}, 094001 (2014).

\bibitem{Wolf08a} M. M. Wolf and J. I. Cirac, Commun. Math. Phys. {\bf 279}, 147 (2008).

\bibitem{Rivas10} \'A. Rivas, S. F. Huelga and M. B. Plenio, Phys. Rev. Lett. {\bf 105}, 050403 (2010). 

\bibitem{Breuer09} H-P. Breuer, E.-M. Laine, and J. Piilo, Phys. Rev. Lett. {\bf 103}, 210401 (2009).

\bibitem{Lu10} X.-M. Lu, X. Wang, and C. P. Sun, Phys. Rev. A {\bf 82}, 042103 (2010). 

\bibitem{Luo12} S. Luo, S. Fu, and H. Song, Phys. Rev. A {\bf 86}, 044101 (2012).

\bibitem{Lorenzo13} S. Lorenzo, F. Plastina, and M. Paternostro, Phys. Rev. A {\bf 88}, 020102(R) (2013).

\bibitem{Liu13} J. Liu, X.-M. Lu, and X. Wang, Phys. Rev. A {\bf 87}, 042103 (2013). 

\bibitem{Bylicka14} B. Bylicka, D.  Chru\'{s}ci\'{n}ski, and S. Maniscalco, Sci. Rep. {\bf 4}, 5720 (2014). 

\bibitem{Chruscinski14a} D. Chru\'{s}ci\'{n}ski and S. Maniscalco, Phys. Rev. Lett. {\bf 112}, 120404 (2014).

\bibitem{LoGullo14} N. Lo Gullo, I. Sinayskiy, Th. Busch, and F. Petruccione, arXiv:1401.1126.

\bibitem{Guarnieri14} G. Guarnieri, A. Smirne, and B. Vacchini, Phys. Rev. A {\bf 90}, 022110 (2014).

\bibitem{Haikka11} P. Haikka, J.D. Cresser, and S. Maniscalco, Phys. Rev. A {\bf  83}, 012112 (2011). 

\bibitem{Chruscinski11} D. Chru\'{s}ci\'{n}ski, A. Kossakowski, and A. Rivas, Phys. Rev. A {\bf 83}, 052128 (2011). 

\bibitem{Vacchini11} B. Vacchini, A. Smirne, E.-M. Laine, J. Piilo, and H.-P. Breuer, 
New. J. Phys. {\bf 13}, 093004 (2011).

\bibitem{Jiang13} M. Jiang and S. Luo, Phys. Rev. A {\bf 88}, 034101 (2013).

\bibitem{Wissmann12} S. Wi\ss mann, A. Karlsson, E.-M. Laine, J. Piilo, and H.-P. Breuer, 
Phys. Rev. A {\bf 86}, 062108 (2012).

\bibitem{Hall14} M. J. W. Hall, J. D. Cresser, L. Li, and E. Andersson, Phys. Rev. A {\bf 89}, 042120 (2014). 

\bibitem{Bruning12} E. Br\"uning, H. M\"akel\"a, A. Messina, and F. Petruccione, J. Mod. Opt. {\bf 59}, 1 (2012).

\bibitem{Laine10} E.-M. Laine, J. Piilo, and H.-P. Breuer, Phys. Rev. A {\bf 81}, 062115 (2010). 

\bibitem{He11} Z. He, J. Zou, L. Li, and B. Shao, Phys. Rev. A {\bf 83}, 012108 (2011). 

\bibitem{Makela13} H. M\"akel\"a and M. M\"ott\"onen, Phys. Rev. A {\bf 88}, 052111 (2013). 

\bibitem{Clos12} G. Clos and H.-P. Breuer, Phys. Rev. A {\bf 86}, 012115 (2012). 

\bibitem{Zeng12} H. S. Zeng, N. Tang, Y. P. Zheng, and T. T. Xu, Eur. Phys. J. D {\bf 66}, 255 (2012).

\bibitem{Breuer} H.-P. Breuer and F.~Petruccione, \emph{The Theory of Open Quantum Systems}  (Oxford University Press, Oxford, 2007).

\end{thebibliography}
\end{document}